\documentclass[review,number,sort&compress]{elsarticle} 

\usepackage{subfigure}
\usepackage{amsmath}

\usepackage{lineno}

\def\gluino{\ensuremath{\tilde{g}}}
\def\ninoone{\ensuremath{\mathchoice%
      {\displaystyle\raise.4ex\hbox{$\displaystyle\tilde\chi^0_1$}}%
         {\textstyle\raise.4ex\hbox{$\textstyle\tilde\chi^0_1$}}%
       {\scriptstyle\raise.3ex\hbox{$\scriptstyle\tilde\chi^0_1$}}%
 {\scriptscriptstyle\raise.3ex\hbox{$\scriptscriptstyle\tilde\chi^0_1$}}}}
\def\ninotwo{\ensuremath{\mathchoice%
      {\displaystyle\raise.4ex\hbox{$\displaystyle\tilde\chi^0_2$}}%
         {\textstyle\raise.4ex\hbox{$\textstyle\tilde\chi^0_2$}}%
       {\scriptstyle\raise.3ex\hbox{$\scriptstyle\tilde\chi^0_2$}}%
 {\scriptscriptstyle\raise.3ex\hbox{$\scriptscriptstyle\tilde\chi^0_2$}}}}
\def\chinoonepm{\ensuremath{\mathchoice%
      {\displaystyle\raise.4ex\hbox{$\displaystyle\tilde\chi^\pm_1$}}%
         {\textstyle\raise.4ex\hbox{$\textstyle\tilde\chi^\pm_1$}}%
       {\scriptstyle\raise.3ex\hbox{$\scriptstyle\tilde\chi^\pm_1$}}%
 {\scriptscriptstyle\raise.3ex\hbox{$\scriptscriptstyle\tilde\chi^\pm_1$}}}}

\begin{document}

\begin{frontmatter}
\title{Continuous simulation of hypothetical physics processes with multiple free parameters}

\author[nju,ipas,oxford]{Jiahang Zhong\corref{cor1}}
\author[nju]{Run-Sheng Huang}
\author[ipas]{Shih-Chang Lee}
\cortext[cor1]{Corresponding author. Jiahang.Zhong@cern.ch, (+41)0227674478. B104 2-C07 CERN, CH-1211, Switzerland.}
\address[nju]{School of Physics, Nanjing University, Nanjing, CN - Jiangsu 210093, China}
\address[ipas]{Institute of Physics, Academia Sinica, TW - Taipei 11529, Taiwan}
\address[oxford]{Department of Physics, University of Oxford, UK - Oxford OX13PU, United Kingdom}

\begin{abstract}
We propose a new approach to simulate hypothetical physics processes which are defined by multiple free parameters. Compared to the conventional grid-scan approach, the new method can produce accurate estimations of the detector acceptance and signal event yields continuously over the parameter space with fewer simulation events. The performance of this method is illustrated with two realistic cases.
\end{abstract}

\end{frontmatter}

\section{Introduction}
\label{sec:Intro}

As data collected by experiments at the Large Hadron Collider (LHC) is rapidly increasing, many efforts have been devoted to searches for various Beyond the Standard Model (BSM) physics at the energy frontier. Many BSM models are defined by multiple free parameters, such as the masses of hypothetical particles and their coupling constants, which are to be constrained by data. The various cross sections and branching ratios, as well as detector acceptance and signal event yields, depend on the parameters. Consequently, the search results for BSM physics are usually expressed in terms of excluded regions in the multi-dimensional parameter space.

In most BSM searches, the hypothetical signal processes are studied in the ``grid-scan" approach, i.e. Monte-Carlo (MC) samples are simulated with specific values of the parameters, corresponding to discrete points on a grid in the parameter space. Each MC sample generally contains a considerable number of events for adequate statistical precision. The signal yields are estimated from MC results of each sample and interpolated between these grid points. However, to produce many samples with full-detector simulation is computationally expensive, especially if sufficient granularity in the multivariate parameter space is desired. It suffers from the so-called ``curse of dimensionality", i.e. the number of grid points needed increases exponentially with the number of free parameters. Usually, the grid-scanned parameter space is limited to two dimensions at a time, keeping the other parameters fixed. Moreover, the grid is often constructed sparsely. As a result, the search limits derived are subject to systematic uncertainties from the interpolation which have to be estimated.

In this paper, we propose a new method to achieve continuous prediction of signal yields over the entire parameter space, under the assumption that the acceptance is a smooth function of the parameters. With such an assumption, the simulated events on neighboring space points are no longer treated independently, therefore fewer events are needed in total. 

This paper is organized as follows. In section~\ref{sec:Method} we describe the technical procedure of this method. Then we illustrate the performance of the approach in section~\ref{sec:LRSM}, with a 2-D example which simulates the production of the hypothetical right-handed W boson and heavy Majorana neutrino in the Left-Right Symmetric Model (LRSM). Another 4-D example of a simplified SuperSymmetric (SUSY) model is demonstrated in section~\ref{sec:TwoStepsGluino}, to highlight the advantage in higher-dimensional parameter space. In section~\ref{sec:Discussion}, we discuss potential improvements that can be implemented in various special scenarios. 

\section{Continuous simulation}
\label{sec:Method}
In order to estimate the signal yield, two parts of information are needed from MC simulation. One is the production rate of the considered process (denoted as $\sigma$), which includes the cross-section and the branching ratio. This can normally be obtained at the event generation level and is relatively light-weight from a computational point of view. The more relevant part is the detector acceptance $\epsilon$, namely the probability for the events to be successfully reconstructed and selected. This is related to the detector coverage and the reconstruction efficiency, for which many MC events with parton showers and full detector-simulation are needed for a reliable estimation. The signal yield (N) is the product of these two components and the integrated luminosity ($\mathcal{L}$), 
\begin{equation}
\label{eqn:nYield}
\begin{split}
&N=\sigma\epsilon\mathcal{L} \\
\end{split}
\end{equation}

Considering the fact that detector simulation is the dominant component of the computing requirement, it is crucial to reduce the statistics demand of the acceptance study. In the grid-scan method, this is achieved by limiting the grid size in the parameter space and correspondingly the number of MC samples. However, a considerable number of events are still required within each sample, in order to evaluate the acceptance $\epsilon$ with sufficient statistical accuracy. 

Contrary to the grid-scan method, the continuous simulation technique simulates events over more parameter space points, either stochastically distributed or on a grid of very fine granularity. The overall number of MC events needed is reduced by generating only a few events at each selected parameter space point. If we treated each point independently, little statistical precision on the determination of $\epsilon$ could be achieved with those few events alone. However, under the assumption that $\epsilon$ changes smoothly with respect to the parameters $\mathbf{x}$, the MC events on neighboring points help to constrain $\epsilon$. This is realized by fitting the function $\epsilon(\mathbf{x})$ using a Bayesian Neural Network (BNN) ~\cite{Mackay:1995}.

Neural Network (NN) is a very powerful technique for multivariate pattern recognition as it can handle multi-dimensional non-linear data without suffering much of the "curse of dimensionality". In the past decade, it has been extensively used in the field of Particle Physics. In most applications it was used to construct a multivariate discriminant, with the aim of separating signals from backgrounds~\cite{Abazov:2001ns,Chiappetta:1993zv,Abramowicz:1995zi}. On the other hand, NN has long been recognized as a universal approximator for multivariate functions, as long as it has sufficient neurons~\cite{Cybenko:1989,Hornik1989359}. This functionality as a non-parametric regression tool has also been exploited by physicists in recent years~\cite{NNPDF,Joan2006,Graczyk:2010gw}. 

Our usage of BNN to fit the acceptance distribution is also a regression application. It differs from previous regression applications in that the output of our function $\epsilon(\mathbf{x})$ is a probability taking values between 0 and 1, and that our input samples are MC events with Boolean target values to indicate whether the event is selected or not. The grid-scan approach needs to accumulate many MC events at each single point before it can evaluate $\epsilon$ to a good precision. Instead, the BNN algorithm we used here~\cite{jzhong2011} estimates the whole probability distribution in an unbinned manner. It can achieve similar precision to the grid-scan method with much fewer MC events.

This unbinned regression is carried out as follows. For each generated MC event, the corresponding parameters of the model $\mathbf{x}$ are passed to the BNN as the input variables. In addition, a target value $t$ of 1 or 0 is assigned to each event, depending on whether it survived the selection cuts. Then the BNN is fitted (or "trained") to have its output $y(\mathbf{x})$ approximating the probability distribution $\epsilon(\mathbf{x})$ by minimizing the cross-entropy cost function 
\begin{equation} 
\label{eqn:CE}
\begin{split}
\text{CE}&=\sum_k (-t_k\log y(\mathbf{x_k})-(1-t_k)\log(1-y(\mathbf{x_k})))
\end{split}
\end{equation}
where $k$ loops over all simulated events. The cross-entropy function is derived from Bernoulli likelihood and naturally attributes the meaning of success probability to the output, $y$. 

The trained BNN can then be used to estimate $\epsilon(\mathbf{x})$. Given any point $\mathbf{x}$ in the parameter space, the BNN will output the estimated value of the acceptance. Multiplying it to the production rate $\sigma(\mathbf{x})$ estimated from MC generators and the integrated luminosity, we obtain the corresponding signal yield at any $\mathbf{x}$.

The Bayesian implementation of the NN provides not only the fitted value but also the estimation of the uncertainty for each predicted $\epsilon(\mathbf{x})$~\cite{jzhong2011}. This uncertainty estimation reflects the statistical fluctuation of the input events as well as the goodness of BNN fitting. It is based on the same training sample so that no additional simulation and computing resources are needed.

\section{Demo: 2-D Left-Right Symmetric Model}
\label{sec:LRSM}

\begin{figure}[htbp]
\begin{center}
  \includegraphics[width=0.4\textwidth]{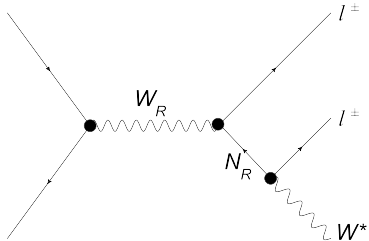}
\caption{The Feynman diagram of $W_R$ production in the LRSM model. The right-handed W boson couples with a lepton and a hypothetical Majorana neutrino, which subsequently decays into another lepton and two jets.}
\label{fig:FD_WRNR}
\end{center}
\end{figure}

In the following we will demonstrate the application of the continuous MC technique by simulating the hypothetical production of a right-handed W boson ($W_R$) which subsequently decays into a heavy Majorana neutrino ($N_R$) as predicted in the Left-Right Symmetrical Models (LRSM).  This is a typical BSM model searched for at the LHC~\cite{LRSM}. Two leptons are expected in the final state, as shown in figure~\ref{fig:FD_WRNR}.
	
The masses of the two hypothetical particles, $M(W_R)$ and $M(N_R)$, are generally considered as the free parameters of this model. These values affect not only the cross-section of the process, but also the kinematics used in the event selection. In this example, we considered the parameter space region with $W_R$ mass between 0.5 TeV and 1.5 TeV, which is being actively studied at the LHC experiments. The $N_R$ mass was limited to be smaller than the $W_R$ mass in order to accommodate the concerned decay mode.

The Pythia generator~\cite{Pythia} is used to simulate this process. For simplicity, detector effects are not included in this demonstration. Only the acceptance $\epsilon(\mathbf{x})$ at the generator level is studied. From the above-mentioned mass ranges we picked 100,000 points stochastically, which are dense enough to cover the 2D parameter space. Then 10 MC events were generated at each point so that we have a total of 1 million events simulated for this demonstration of continuous MC.  The reason for generating multiple events at each point is to reduce the overhead of generator re-initialization and the computing load of BNN training. However, we still maintain a relatively low number of events at each point in order to spread over as many space points as possible.

The event selection we assumed requires at least two leptons ($e$,$\mu$) with transverse momentum greater than 20 GeV and pseudo-rapidity $\eta$ between -2.5 and 2.5. A 100\% fiducial efficiency is assumed for lepton identification. In addition, a common isolation cut for background suppression is applied, requiring the transverse energy flow of all interactive particles inside a cone of size  $\Delta R=\sqrt{(\Delta\eta)^2+(\Delta\phi)^2}=0.2$ around the lepton to be less than 10\% of the lepton's transverse momentum. 

\begin{figure}[htbp]
\begin{center}
  \subfigure[]{
     \label{fig:LRSMBNN}
     \includegraphics[width=0.7\textwidth]{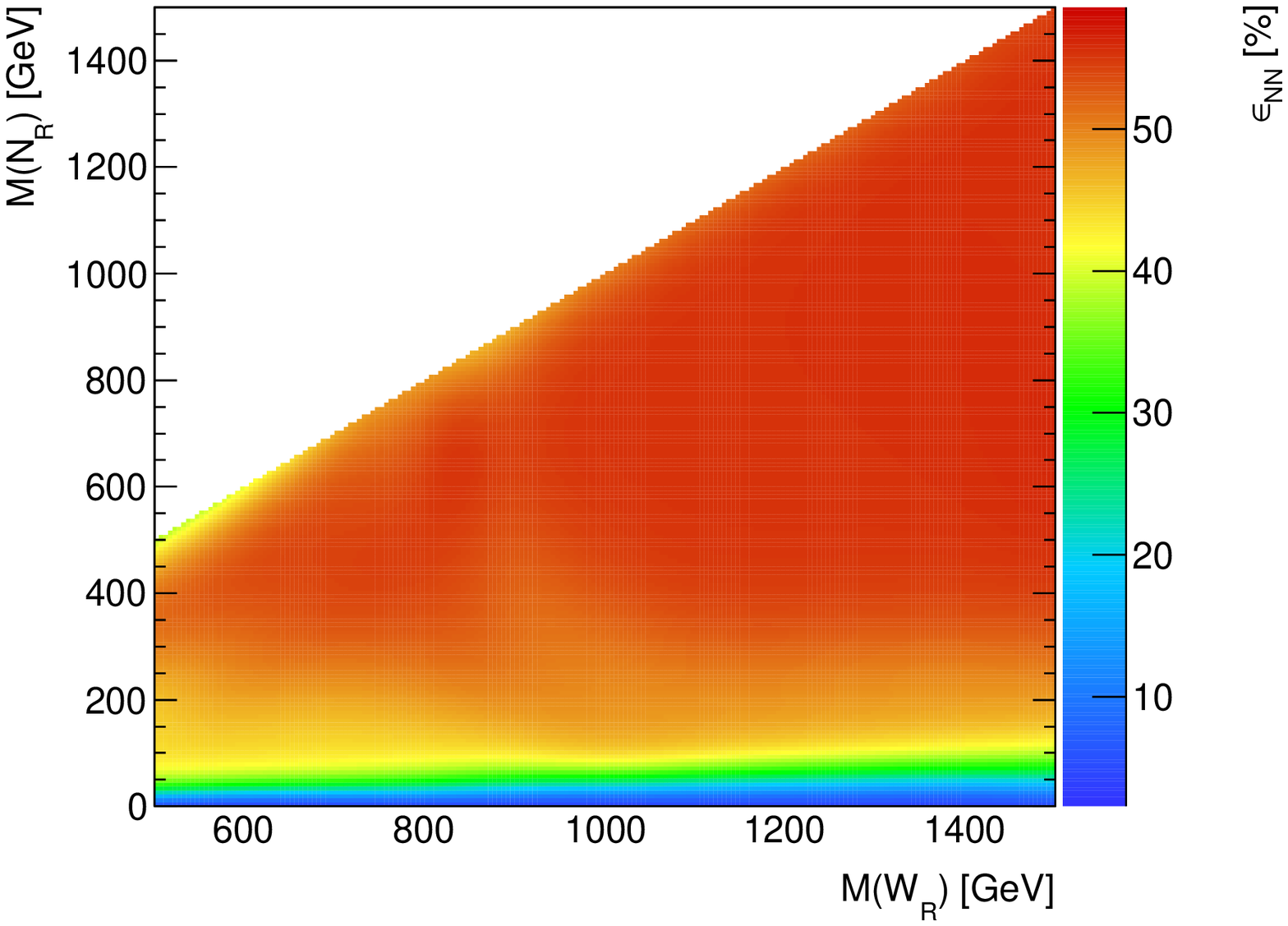}
  }
  \subfigure[]{
     \label{fig:LRSMGS}
     \includegraphics[width=0.7\textwidth]{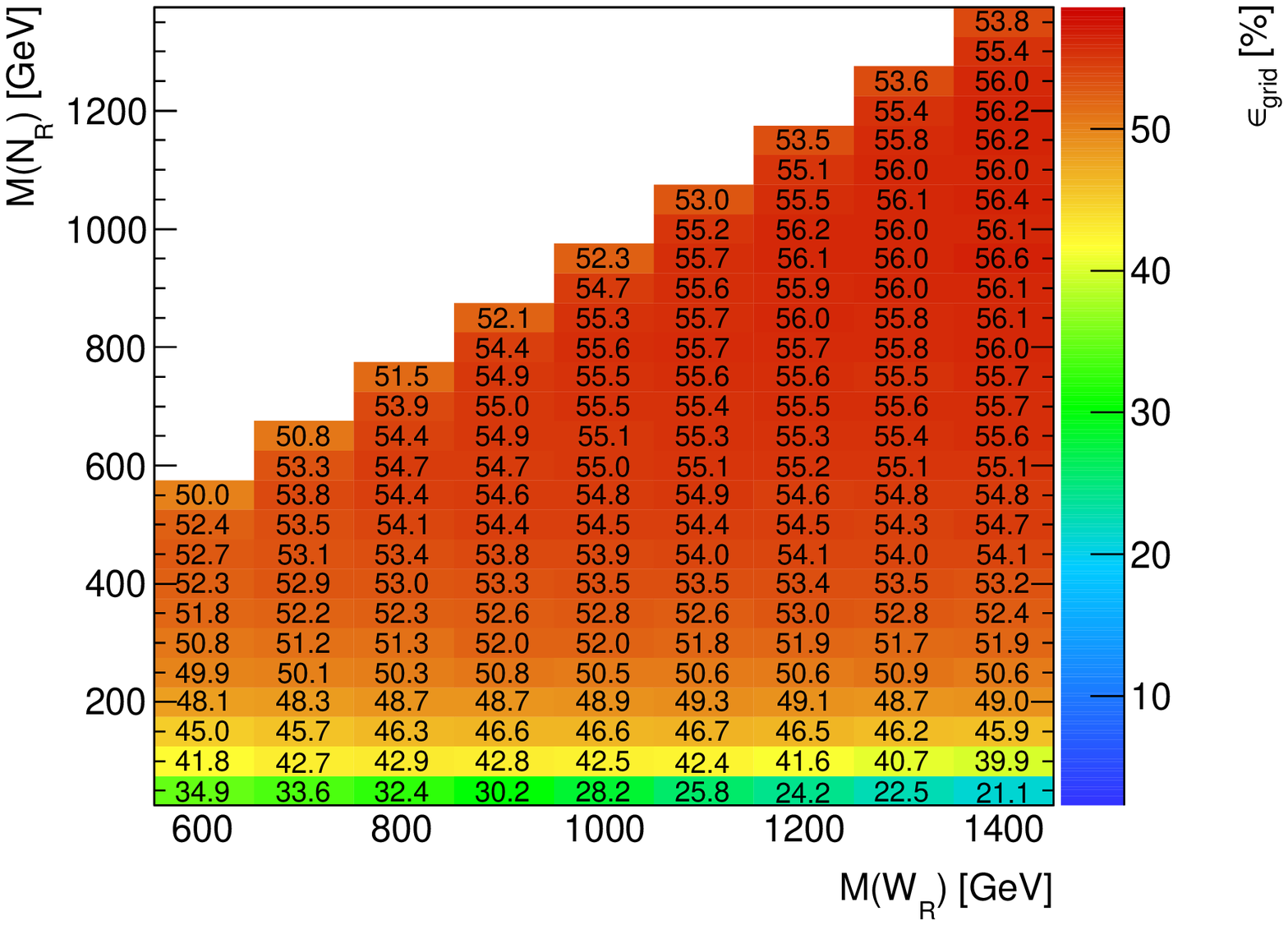}
  }
\caption{The acceptance distribution fitted by BNN (a) and estimated by the grid-scan samples (b).}
\label{fig:LRSMDist}
\end{center}
\end{figure}

The acceptance of this event selection is fitted by a BNN with forty neurons in the first hidden layer and ten neurons in the second hidden layer. This topology is arbitrarily chosen, merely to provide redundant complexity within the computing budget. The BNN can automatically constrain the complexity and avoid over-fitting by a regulator mechanism, which is tuned based on the Bayesian evidence framework~\cite{Mackay:1995,jzhong2011}. The fitted distribution of $\epsilon(\mathbf{x})$ is given in figure~\ref{fig:LRSMBNN}. 

To test the performance, we also simulated another set of discrete samples by the grid-scan approach. The values of $M(W_R)$ are from 600 GeV to 1400 GeV with 100 GeV steps, and the values of $M(N_R)$ are from 50 GeV to $(M(W_R)-50)$ GeV with 50 GeV steps. A total of 171 samples have been simulated with these points, each containing 100,000 events for statistical accuracy. The acceptance function evaluated from these samples is shown in figure~\ref{fig:LRSMGS}.  

\begin{figure}[htbp]
\begin{center}
  \subfigure[]{
     \label{fig:LRSMDif_2D}
     \includegraphics[width=0.7\textwidth]{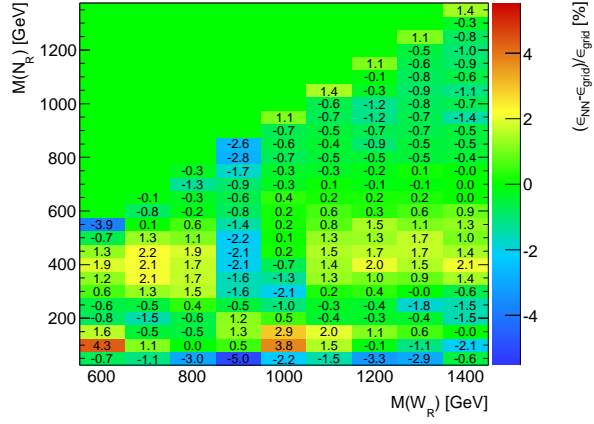}
  }
  \subfigure[]{
     \label{fig:LRSMDif_1D}
     \includegraphics[width=0.7\textwidth]{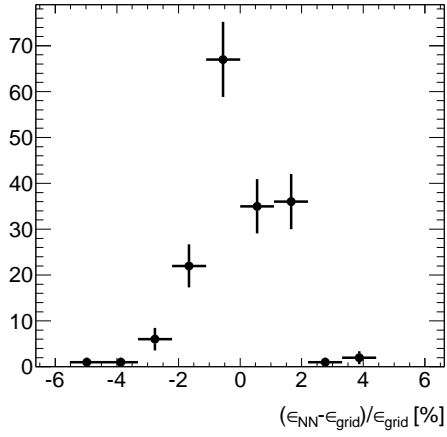}
  }
\caption{The relative deviation of the BNN predictions from the estimations by grid-scan, at the selected parameter space points (a), and the distribution of these deviations (b).}
\label{fig:LRSMDif}
\end{center}
\end{figure}

We then compared the $\epsilon(\mathbf{x})$ values fitted by the BNN at these grid points to the reference values directly estimated from the discrete samples. The relative deviations are shown in figure~\ref{fig:LRSMDif}. For most points, the deviation is well within a few percent of the reference value, which is negligible compared to the systematic uncertainties of simulation. Notably, the relative deviation becomes larger when $M(N_R)$ approaches to 0. This is partly due to the smaller value of $\epsilon(\mathbf{x})$ itself in this region. In addition, when approaching the edge of the parameter space of the training sample, the lack of neighboring points increases the uncertainty of the fitting.
      
\begin{figure}[htbp]
\begin{center}
  \subfigure[]{
     \label{fig:LRSMDoR_2D}
     \includegraphics[width=0.7\textwidth]{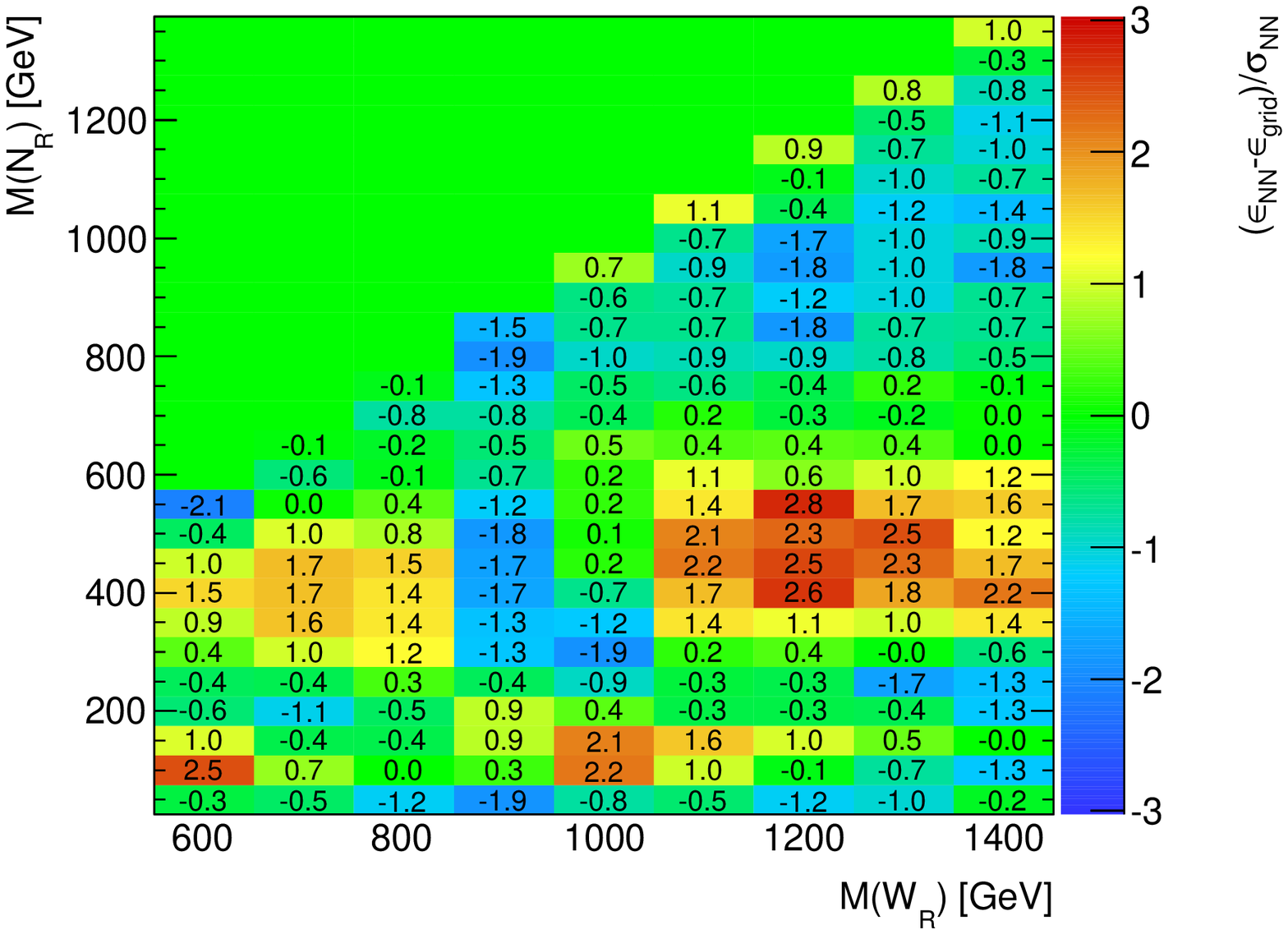}
  }
  \subfigure[]{
     \label{fig:LRSMDoR_1D}
     \includegraphics[width=0.7\textwidth]{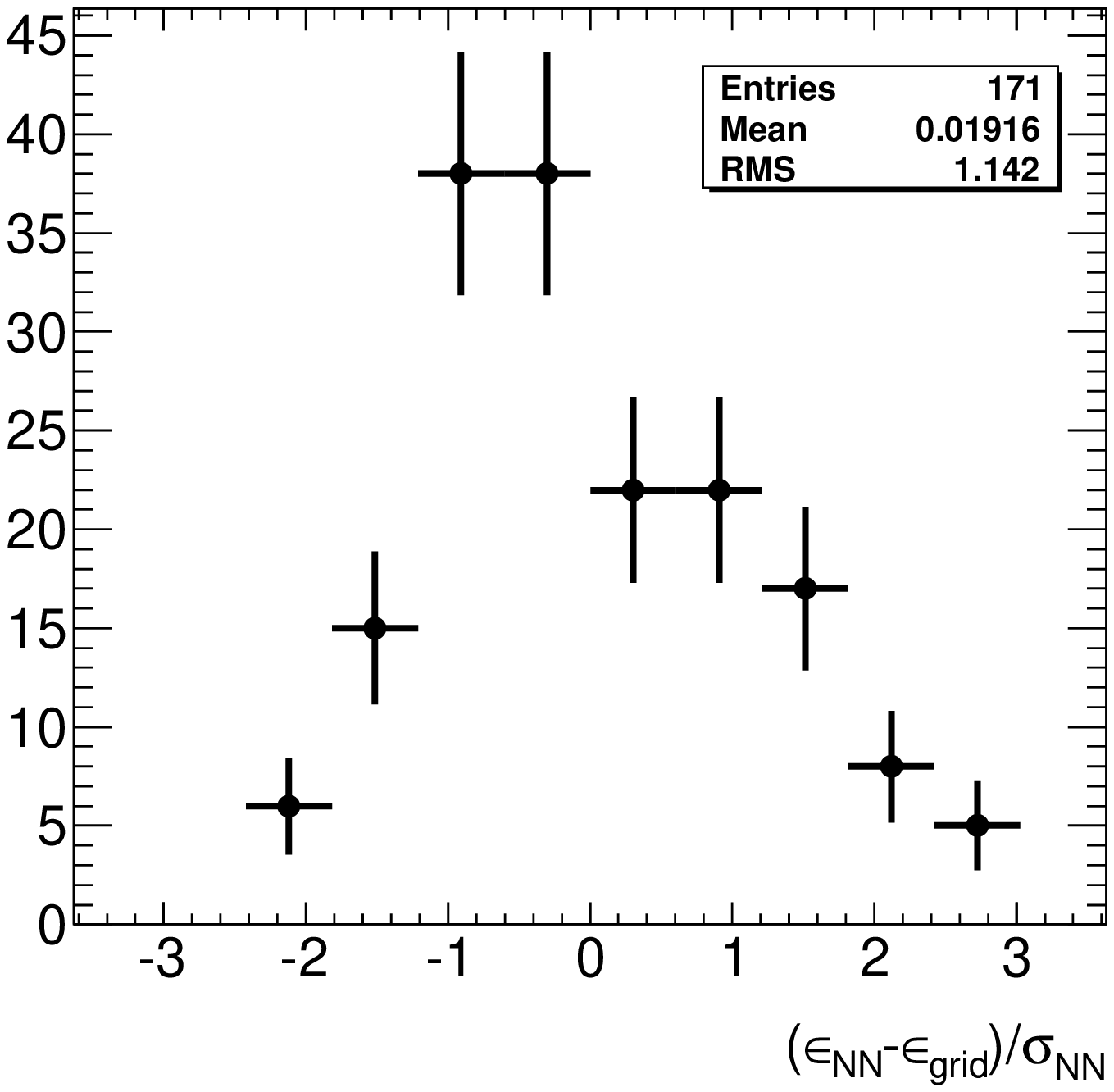}
  }
\caption{The deviations between the BNN predictions and the estimations by grid-scan, divided by the BNN uncertainties, at the selected parameter space points (a), and the distribution of these ratios (b).}
\label{fig:LRSMDoR}
\end{center}
\end{figure}

As mentioned in section~\ref{sec:Method}, the BNN also estimates the uncertainties of the fitted probabilities, denoted as $\sigma_{NN}(\mathbf{x})$. The uncertainties account for the deviation we observed, both in the central region with stable $\epsilon(\mathbf{x})$ and at the edge region where $M(N_R)$ approaches 0. By comparing the deviation between the fitted value and the reference value to the BNN estimated uncertainty (figure~\ref{fig:LRSMDoR}), we can see good consistency over the entire parameter space.
 
Only 1 million MC events were used by our example of continuous simulation. For the grid scan technique, a total of 17.1 million events at 171 discrete points were used to build a typical set of samples in the same parameter space region. To quantitatively compare the statistics demand, we calculated the number of events at each grid point that would yield a statistical uncertainty equal to the BNN uncertainty at the same point. It is worth mentioning that the BNN uncertainty includes not only the statistical fluctuation, but also the uncertainty on the goodness of fitting. This comparison shows, ignoring the interpolation errors, a total number of at least 6.3 million events for these 171 grid-scan points are needed in order to reach a similar precision as the 1 million continuous MC samples do.

\begin{figure}[htbp]
\begin{center}
  \subfigure[]{
     \label{fig:LRSMItpX}
     \includegraphics[width=0.7\textwidth]{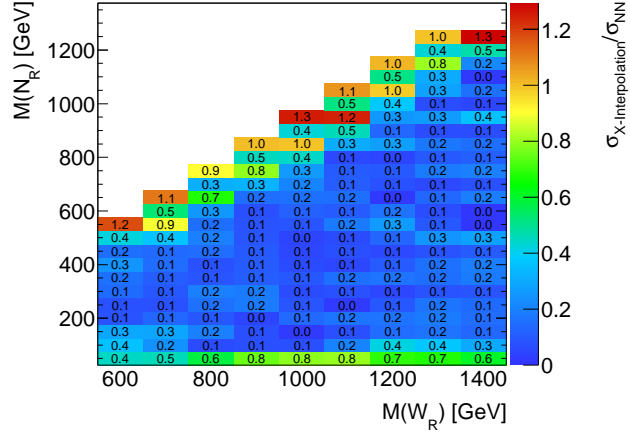}
  }
  \subfigure[]{
     \label{fig:LRSMItpY}
     \includegraphics[width=0.7\textwidth]{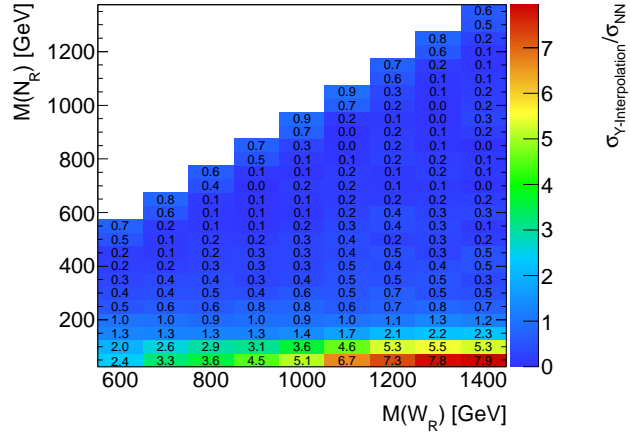}
  }
\caption{The interpolation uncertainties evaluated by comparing to nearby points, in the directions of M($W_R$) (a) and M($N_R$) (b) respectively.}
\label{fig:LRSMItp}
\end{center}
\end{figure}

As mentioned above, the BNN uncertainty includes the uncertainty on the goodness of fitting. The counterpart in the grid-scan approach is the interpolation uncertainty when estimating the acceptance at any point not on the grid. To evaluate this uncertainty for the grid-scan method, we assumed that the value at the interpolated point is uniformly distributed between the values at the neighboring grid points in each dimension. The uncertainty is estimated as the standard deviation of this distribution, i.e. $1/\sqrt{12}$ of the difference between the two values. We averaged the upward and downward uncertainties (if they exist) to get an uncertainty for each grid point. This uncertainty of grid-scan is compared to the BNN uncertainty of continuous simulation, as shown in figure~\ref{fig:LRSMItp}. We can see that the interpolation could often introduce a non-negligible uncertainty, especially in the low $N_R$ mass region where the acceptance changes steeply. To achieve similar interpolation precision as the BNN fitting, the grid-scan will need finer granularity and more statistics, as well as extra effort for optimization of the grid structure.

\section{Demo: 4-D Simplified SuperSymmetric Model}
\label{sec:TwoStepsGluino}

In this section, we will illustrate the performance of the continuous simulation in a 4-D scenario. This is gluino pair production with a two-step cascade decay in one of the Simplified SuperSymmetric Models~\cite{PhysRevD.79.075020,Alves:2011wf}. In this model, the gluino (\gluino) can decay to the Lightest Supersymmetric Particle (LSP), assumed to be the \ninoone{}, through the decay chain of \ensuremath{\gluino \rightarrow qq\chinoonepm \rightarrow qqW^{\pm}\ninotwo \rightarrow qqW^{\pm}Z\ninoone}. The BSM parameters in this model are the masses of the four SUSY particles. For this demonstration, we studied the mass range of \gluino{} between 200 GeV and 1200 GeV. To accommodate the decay chain concerned, the mass of the \chinoonepm{} was limited to between 200 GeV and M(\gluino), and the mass of the \ninotwo{} was limited to between 100 GeV and M(\chinoonepm). Finally, the mass of the LSP was limited to between 0 GeV and M(\ninotwo).

The Herwig++ generator~\cite{Herwigpp} was used to simulate this process. For the continuous MC sample, a total number of 1 million events were generated at 100k space points, stochastically distributed in the parameter space region. For simplicity, the study is again limited to generator level acceptance. We looked for final states with at least two leptons ($e$,$\mu$) with transverse momentum greater than 20 GeV and with pseudo-rapidity $\eta$ between -2.5 and 2.5. A 100\% fiducial efficiency was assumed for lepton identification. In addition, the transverse energy flow of all interactive particles around the lepton within a $\Delta R=0.4$ cone was required to be below 5 GeV. As a typical SUSY signature, we also required the visible energy imbalance in the transverse direction to be greater than 50 GeV. The acceptance of this di-lepton plus missing transverse energy selection was then parameterized by a BNN of the same topology as in the previous example.

For comparison, we also simulated discrete samples by grid-scan, with 100 GeV steps in all four directions. The values of M(\gluino) are from 250 GeV up to 1150 GeV. The values of M(\chinoonepm) are from 200 GeV to M(\gluino)-50 GeV. The values of M(\ninotwo) are from 100 GeV to M(\chinoonepm)-100 GeV. And the values of  M(LSP) are from 0 GeV to M(\ninotwo)-100 GeV. These made up a 4-D grid with 715 points, with 30k events simulated at each point. 

\begin{figure}[htbp]
\begin{center}
  \includegraphics[width=0.7\textwidth]{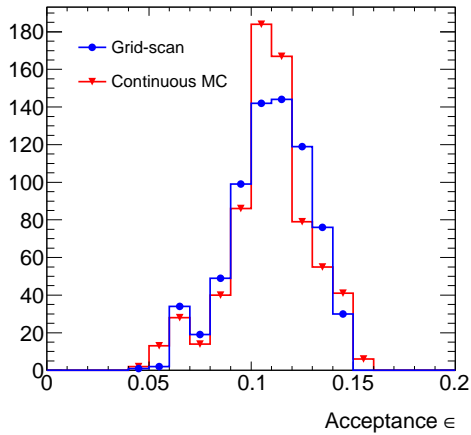}
\caption{The acceptances at 715 grid points, evaluated from the grid-scan samples and the BNN fitting.}
\label{fig:SUSYeff}
\end{center}
\end{figure}

\begin{figure}[htbp]
\begin{center}
  \includegraphics[width=0.7\textwidth]{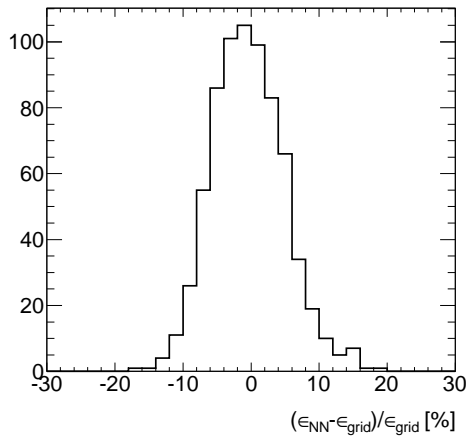}
\caption{The relative deviation of BNN estimations from the grid-scan reference values.}
\label{fig:SUSYdif}
\end{center}
\end{figure}

With these 715 grid-scan samples, we obtained reference values of $\epsilon(\mathbf{x})$ ranging from 0.04 to 0.16. Using the BNN trained by the continuous MC sample, we observed similar distributions of $\epsilon(\mathbf{x})$, as shown in figure~\ref{fig:SUSYeff}. Further comparison of the BNN estimations and the reference values at each point revealed that most of them are consistent, well within 10\% relative deviation (fig.~\ref{fig:SUSYdif}). 

\begin{figure}[htbp]
\begin{center}
  \includegraphics[width=0.7\textwidth]{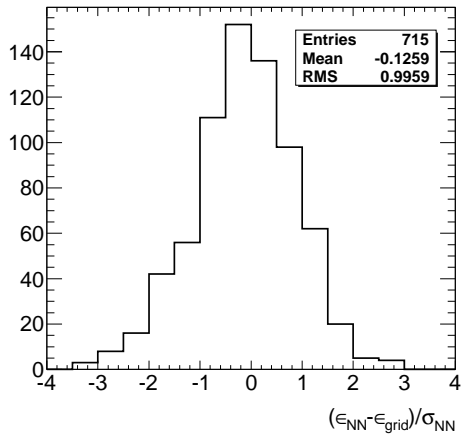}
\caption{The deviation between the BNN estimation and the grid-scan reference value, divided by the BNN uncertainties.}
\label{fig:SUSYdor}
\end{center}
\end{figure}

We also examined the uncertainty estimated by the BNN, comparing it to the actual difference between BNN estimation and the reference value. From figure~\ref{fig:SUSYdor}, we can see that the uncertainties are certainly consistent with the actual deviation.

To demonstrate the statistical efficiency of continuous simulation, we again calculated the minimal number of events at each grid point, if statistical precision equivalent to the BNN uncertainty is desired. The comparison shows that, ignoring the interpolation errors, at least 3.2 million events on the 715 considered grid points are needed, which is already greater than the 1 million statistics we used for the continuous MC sample. 

\begin{figure}[htbp]
\begin{center}
  \subfigure[]{
     \label{fig:SUSYitps}
     \includegraphics[width=0.7\textwidth]{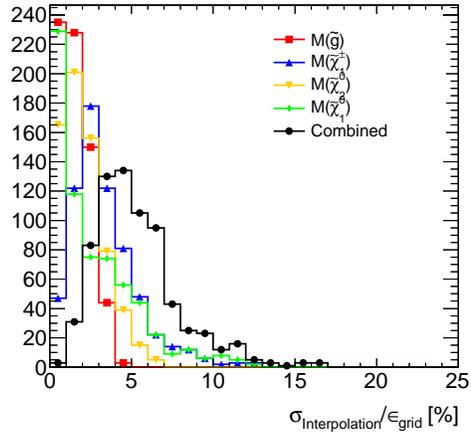}
  }
  \subfigure[]{
     \label{fig:SUSYitpsR}
     \includegraphics[width=0.7\textwidth]{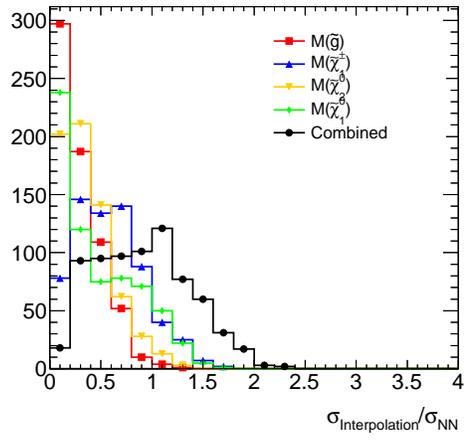}
  }
\caption{The interpolation uncertainties on all fours axes and their quadratic combination, comparing to the reference value of acceptance (a) and the BNN uncertainty (b).}
\label{fig:SUSYitp}
\end{center}
\end{figure}

The grid-scan suffers more uncertainties of interpolation in a 4-D space. With the same measurements as in section~\ref{sec:LRSM}, we estimated the interpolation uncertainties on each of the four parameters as well as their quadratic sum (fig.~\ref{fig:SUSYitps}). Notably, in each direction, due to the constraint of the mass hierarchy, 55 out of the 715 points cannot have this interpolation uncertainty measured, because they have no neighboring points. Overall there are 189 points whose interpolation uncertainties are underestimated due to this. Nevertheless, comparing them with the BNN uncertainties, we can see that the interpolation errors are worse than the BNN uncertainties at a large fraction of the parameter space points (fig.~\ref{fig:SUSYitpsR}). This indicates the necessity of a grid with finer granularity. However, increasing the number of points in each dimension by a mere factor of two would already mean 16 times larger statistics: a manifestation of the effect of the "curse of dimensionality".

\section{Discussion}
\label{sec:Discussion}
Several potential improvements beyond what has been demonstrated can be made depending on the subject under study. For example, the choice of parameter space points can be more intelligently informed. In our examples, the parameters $\mathbf{x}$ are chosen from the space with a flat probability distribution. However, after the first round of acceptance fitting, we may identify certain parameter space regions in which the fitted values have relatively larger uncertainties, either due to the smaller values of $\epsilon(\mathbf{x})$ or due to the proximity to the edge of the parameter space of the training samples. A typical case is the low $M(N_R)$ region in LRSM production, as shown in section~\ref{sec:LRSM}. If enhancement of precision for these regions is desired, it can be achieved by simulating additional events in these regions. Fitting another BNN with the enlarged sample will then give more accurate result for such regions. 

In case it is impractical to produce additional MC events, an alternative solution is to fit another BNN with the existing sample, but assign higher weights during the fitting to those events in the concerned regions. This is equivalent to repeatedly using these events as additional samples. Although the statistical fluctuation remains, it will drive the fitting to focus more on the numerical feature of these regions and rely less on the interpolation from other regions. 

Another potential improvement is to split the parameter space according to different physics processes. The BNN fitting assumes smooth distribution over the parameter space. However, this assumption of smoothness may not hold at boundaries where the physics is different on each side. If such boundaries are known as {\it a priori} knowledge, independent BNNs should be applied on each sub-set of the parameter space within which the signal yield is smooth.

Besides fitting the acceptance $\epsilon(\mathbf{x})$, the production rate $\sigma(\mathbf{x})$ can also be fitted by the BNN. This might be necessary if a continuous function of $\sigma(\mathbf{x})$ is desired or if calculating cross-sections for a large number of parameter space points is impractical. The way of BNN fitting needs to be changed slightly for such a regression application. The NN output $y(\mathbf{x})$ should be allowed to range over all real numbers, while the target value will be the $\sigma(\mathbf{x_i})$ for each event. The cost function will be the Mean-Square-Error between $y(\mathbf{x_i})$ and $\sigma(\mathbf{x_i})$, representing a Gaussian likelihood. As a result, the signal expectation $N(\mathbf{x})$ can also be obtained continuously over the parameter space.

\section{Conclusion}
\label{sec:Conclusion}
In this work, we introduced a new approach to simulate hypothetical processes with multiple free parameters and its use in estimating signal acceptance and yields continuously over the parameter space. In particular, the BNN technique is used for unbinned fitting of the acceptance, which is the key to providing continuous estimation and reducing the required number of simulation events. The two examples we showed both suggest that the acceptance can be well estimated with fewer MC events than needed in the grid-scan approach. In addition, the uncertainty of the fitted signal yield is estimated by the BNN simultaneously which is a natural consequence of the Bayesian inference.      

\section*{Acknowledgements}

We thank Song-Ming Wang, Rachid Mazini, Sarah Livermore for helpful discussions. Jiahang Zhong and Shih-Chang Lee are partially supported by the National Science Council, Taiwan under the contract number NSC99-2119-M-001-015. Jiahang Zhong is also supported by the UK Science and Technology Facilities Council.

\bibliographystyle{model1a-num-names}
\bibliography{ContinuousMC}

\begin{thebibliography}{15}
\expandafter\ifx\csname natexlab\endcsname\relax\def\natexlab#1{#1}\fi
\providecommand{\bibinfo}[2]{#2}
\ifx\xfnm\relax \def\xfnm[#1]{\unskip,\space#1}\fi
\bibitem[{MacKay(1995)}]{Mackay:1995}
\bibinfo{author}{D.~J.~C. MacKay}, \bibinfo{journal}{Network: Computation in
  Neural Systems} \bibinfo{volume}{6} (\bibinfo{year}{1995})
  \bibinfo{pages}{469--505}.
\bibitem[{Abazov et~al.(2001)}]{Abazov:2001ns}
\bibinfo{author}{V.~M. Abazov}, et~al., \bibinfo{journal}{Phys. Lett.}
  \bibinfo{volume}{B517} (\bibinfo{year}{2001}) \bibinfo{pages}{282--294}.
\bibitem[{Chiappetta et~al.(1994)Chiappetta, Colangelo, De~Felice, Nardulli,
  and Pasquariello}]{Chiappetta:1993zv}
\bibinfo{author}{P.~Chiappetta}, \bibinfo{author}{P.~Colangelo},
  \bibinfo{author}{P.~De~Felice}, \bibinfo{author}{G.~Nardulli},
  \bibinfo{author}{G.~Pasquariello}, \bibinfo{journal}{Phys. Lett.}
  \bibinfo{volume}{B322} (\bibinfo{year}{1994}) \bibinfo{pages}{219--223}.
\bibitem[{Abramowicz et~al.(1995)Abramowicz, Caldwell, and
  Sinkus}]{Abramowicz:1995zi}
\bibinfo{author}{H.~Abramowicz}, \bibinfo{author}{A.~Caldwell},
  \bibinfo{author}{R.~Sinkus}, \bibinfo{journal}{Nucl. Instrum. Meth.}
  \bibinfo{volume}{A365} (\bibinfo{year}{1995}) \bibinfo{pages}{508--517}.
\bibitem[{Cybenko(1989)}]{Cybenko:1989}
\bibinfo{author}{G.~Cybenko}, \bibinfo{journal}{Mathematics of Control,
  Signals, and Systems (MCSS)} \bibinfo{volume}{2} (\bibinfo{year}{1989})
  \bibinfo{pages}{303--314}.
\bibitem[{Hornik et~al.(1989)Hornik, Stinchcombe, and White}]{Hornik1989359}
\bibinfo{author}{K.~Hornik}, \bibinfo{author}{M.~Stinchcombe},
  \bibinfo{author}{H.~White}, \bibinfo{journal}{Neural Networks}
  \bibinfo{volume}{2} (\bibinfo{year}{1989}) \bibinfo{pages}{359 -- 366}.
\bibitem[{Forte et~al.(2002)Forte, Garrido, Latorre, and Piccione}]{NNPDF}
\bibinfo{author}{S.~Forte}, \bibinfo{author}{L.~Garrido},
  \bibinfo{author}{J.~I. Latorre}, \bibinfo{author}{A.~Piccione},
  \bibinfo{journal}{JHEP} \bibinfo{volume}{05} (\bibinfo{year}{2002})
  \bibinfo{pages}{062}.
\bibitem[{Rojo(2006)}]{Joan2006}
\bibinfo{author}{J.~Rojo}, \bibinfo{journal}{JHEP} \bibinfo{volume}{05}
  (\bibinfo{year}{2006}) \bibinfo{pages}{040}.
\bibitem[{Graczyk et~al.(2010)Graczyk, Plonski, and Sulej}]{Graczyk:2010gw}
\bibinfo{author}{K.~M. Graczyk}, \bibinfo{author}{P.~Plonski},
  \bibinfo{author}{R.~Sulej}, \bibinfo{journal}{JHEP} \bibinfo{volume}{09}
  (\bibinfo{year}{2010}) \bibinfo{pages}{053}.
\bibitem[{Zhong et~al.(2011)Zhong, Huang, and Lee}]{jzhong2011}
\bibinfo{author}{J.~Zhong}, \bibinfo{author}{R.-S. Huang},
  \bibinfo{author}{S.-C. Lee}, \bibinfo{journal}{Computer Physics
  Communications} \bibinfo{volume}{182} (\bibinfo{year}{2011})
  \bibinfo{pages}{2655 -- 2660}.
\bibitem[{Ferrari et~al.(2000)Ferrari, Collot, Andrieux, Belhorma,
  de~Saintignon, Hostachy, Martin, and Wielers}]{LRSM}
\bibinfo{author}{A.~Ferrari}, \bibinfo{author}{J.~Collot},
  \bibinfo{author}{M.-L. Andrieux}, \bibinfo{author}{B.~Belhorma},
  \bibinfo{author}{P.~de~Saintignon}, \bibinfo{author}{J.-Y. Hostachy},
  \bibinfo{author}{P.~Martin}, \bibinfo{author}{M.~Wielers},
  \bibinfo{journal}{Phys. Rev. D} \bibinfo{volume}{62} (\bibinfo{year}{2000})
  \bibinfo{pages}{013001}.
\bibitem[{Sjostrand et~al.(2006)Sjostrand, Mrenna, and Skands}]{Pythia}
\bibinfo{author}{T.~Sjostrand}, \bibinfo{author}{S.~Mrenna},
  \bibinfo{author}{P.~Z. Skands}, \bibinfo{journal}{JHEP} \bibinfo{volume}{05}
  (\bibinfo{year}{2006}) \bibinfo{pages}{026}.
\bibitem[{Alwall et~al.(2009)Alwall, Schuster, and Toro}]{PhysRevD.79.075020}
\bibinfo{author}{J.~Alwall}, \bibinfo{author}{P.~C. Schuster},
  \bibinfo{author}{N.~Toro}, \bibinfo{journal}{Phys. Rev. D}
  \bibinfo{volume}{79} (\bibinfo{year}{2009}) \bibinfo{pages}{075020}.
\bibitem[{Alves et~al.(2011)}]{Alves:2011wf}
\bibinfo{author}{D.~Alves}, et~al., \bibinfo{journal}{ArXiv e-prints
  hep-ph/1105.2838}  (\bibinfo{year}{2011}).
\bibitem[{Bahr et~al.(2008)Bahr, Gieseke, Gigg, Grellscheid, Hamilton,
  Latunde-Dada, Platzer, Richardson, Seymour, Sherstnev, and Webber}]{Herwigpp}
\bibinfo{author}{M.~Bahr}, \bibinfo{author}{S.~Gieseke},
  \bibinfo{author}{M.~Gigg}, \bibinfo{author}{D.~Grellscheid},
  \bibinfo{author}{K.~Hamilton}, \bibinfo{author}{O.~Latunde-Dada},
  \bibinfo{author}{S.~Platzer}, \bibinfo{author}{P.~Richardson},
  \bibinfo{author}{M.~Seymour}, \bibinfo{author}{A.~Sherstnev},
  \bibinfo{author}{B.~Webber}, \bibinfo{journal}{The European Physical Journal
  C - Particles and Fields} \bibinfo{volume}{58} (\bibinfo{year}{2008})
  \bibinfo{pages}{639--707}.

\end{thebibliography}

\end{document}